\documentclass[aps,prl,twocolumn,amsmath,showpacs,amssymb,superscriptaddress,nofootinbib]{revtex4}

\newlength{\picwidth}

\makeatother
\usepackage{graphicx,epstopdf}
\usepackage{dcolumn}
\usepackage{bm}
\usepackage{amssymb,amsmath}
\usepackage{latexsym}
\usepackage{color}
\usepackage[normalem]{ulem}
\usepackage{cancel}
\allowdisplaybreaks
\usepackage{amsfonts}
\usepackage{color}
\usepackage{subfigure}
\newcommand{\be}{\begin{equation}}
\newcommand{\ee}{\end{equation}}

\newcommand{\bes}{\begin{subequations}}
\newcommand{\ees}{\end{subequations}}
\newcommand{\ba}{\begin{eqnarray}}
\newcommand{\ea}{\end{eqnarray}}
\newcommand{\bea}{\begin{eqnarray}}
\newcommand{\eea}{\end{eqnarray}}

\graphicspath{%
     {  }
}

\newcommand{\bma}{\begin{pmatrix}}
\newcommand{\ema}{\end{pmatrix}}

\newcommand{\stell}{\affiliation{ Physics Department, Stellenbosch University,  
7602, South Africa}}
\newcommand{\nithep}{\affiliation{National Institute for Theoretical Physics (NITheP), Bag X1 Matieland, Stellenbosch, 7602, South Africa}}
\newcommand{\Maryland}{\affiliation{Maryland Center for Fundamental Physics \& Joint Space-Science Institute, Department of Physics, University of Maryland, College Park, MD 20742, USA}}

\begin{document}

\title{Orbital resonances around Black holes}

\author{Jeandrew Brink} \nithep \stell
\author{Marisa Geyer} \stell
\author{ Tanja Hinderer} \Maryland 

\date{\today}

\begin{abstract}
We compute the length and timescales associated with resonant orbits in the Kerr Metric for all orbital and spin parameters. Resonance induced effects are potentially observable when the Event Horizon telescope resolves the inner structure of Sgr A*, space-based gravitational wave detectors record phase-shifts in the waveform during the resonant passage of a neutron star as it spirals in to the black hole and in the  frequencies of quasi periodic oscillations for accreting black holes. The onset of geodesic chaos for non-Kerr spacetimes should occur at the resonance locations quantified here.

\end{abstract}
\maketitle

\emph{Introduction.} Resonant phenomena are ubiquitous in multi-frequency systems and are harbingers of the onset of dynamical chaos \cite{2001Natur.410..773M}.
In celestial mechanics, they play an important role in satellite dynamics. Gaps in the asteroid belt 
and the density profile in the rings of Saturn \cite{Saturn,Asteroid} have in large part been sculpted by resonant interactions.
\begin{figure}
\includegraphics[width = 0.97\columnwidth]{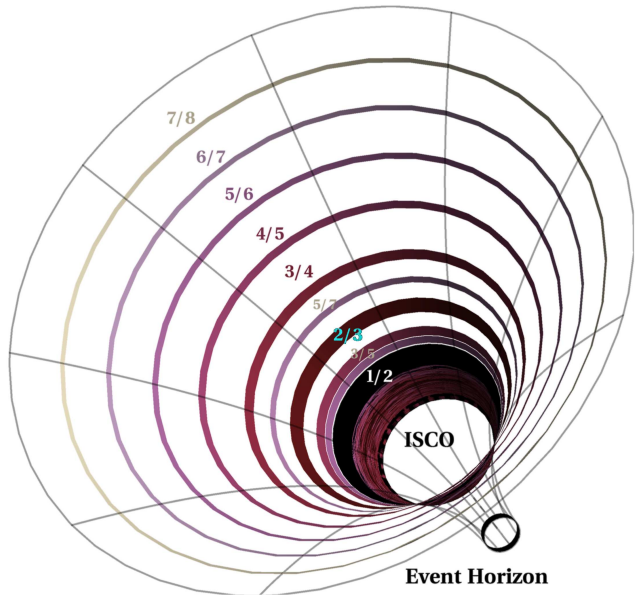}
\caption{\footnotesize{(Color online)
The location of low order resonances around a black hole superimposed on an embedding diagram.  The line width of each resonance is inversely proportional to the order of the resonance  to give an indication of the relative importance of a particular resonance. 
\label{SaturnsRings}}}
\end{figure}
The orbital motion of satellites around black holes is mathematically idealized as bound geodesics in the Kerr metric. Unlike in Newtonian gravity where orbits are characterized by a single rotational frequency $\omega_\phi$, geodesic motion in the Kerr spacetime is described by three frequencies \cite{Schmidt}. The two libration type frequencies 
$\omega_r$ and $\omega_\theta$ corresponding to motion in the radial and longitudinal
directions augment the azimuthal frequency $\omega_\phi$. The existence of these libration type frequencies gives rise to the resonant phenomena considered here.

When exploring dynamics in an astrophysical environment such as near Sgr A* at the Galactic center a number of corrections to the vacuum Kerr Hamiltonian $\mathcal H_K$ must be taken into account. The presence of an accretion disk \cite{Semerák01102012}, other sources of matter, structural deviations of the central black hole away from the Kerr metric \cite{Contopoulos2012, Gair, JdB1},  
the influence of modified gravity, or the mass and
spin of the satellite \cite{lrr-2011-7,spin2} will all affect its orbital motion.
Regardless of the  nature of the perturbation, the  Kolmogorov-Arnold-Moser (KAM) theorem states that the perturbed dynamics 
will be a smooth distortion of Kerr geodesics
provided the frequencies  
of the motion in $\mathcal H_K$ are sufficiently irrational as quantified by the criterion \cite{Arnold1963a, Moser1973}
$
|{m \omega_r - n\omega_\theta} | > K(\epsilon) /(n+m)^{3} \label{eqKAM}
$.
The factor $K(\epsilon)$ here approaches zero as the perturbation vanishes.  
 The notable exception to this theorem is low order ($n+m$ value) resonant orbits whose frequencies occur in the rational ratios $\omega_r/\omega_\theta=n/m=$ $1/2$, $1/3$, $2/3$, $ \cdots$.
  For these orbits the possibility of dramatic deviations from  Kerr dynamics exists. Since the predictions of the KAM theorem depend on $\mathcal H_K$ only we expect a measurable imprint of Kerr's resonant structure in any astrophysical environment. 
The relative locations of low order resonances is shown in the embedding diagram in Fig.~\ref{SaturnsRings} and the characteristic time and length scales associated with each resonance tabulated in Table~\ref{TabRes}. For Sgr A*, the typical timescales for the low-order resonances are $\sim 1$hr 
and the resonances occur at distances of $\sim 50\mu {\rm as}$ 
from the black hole.

Within the next ten years radio telescopes will attain sufficient sensitivity to resolve length scales typical of resonant phenomena at the center of our 
galaxy. The event horizon telescope is one such  observational effort currently under development \cite{eventhorizontel}  using the VLBI network. 
As a neutron star or stellar mass black hole spirals into a super-massive black hole,
it will sample all the resonant bands depicted in Fig.~\ref{SaturnsRings}. Future space based gravitational wave detectors 
may observe resonance-induced shifts in the phasing of the emitted gravitational waves \cite{prl, uchopol}. X-ray, optical and infrared telescopes do not have the resolving power to image Sgr A* directly but can potentially record flux variations from this region that may display timescales characteristic of resonant events \cite{2011ApJ...727L..36F}. 
Quasi-periodic oscillations (QPOs) observed in the X-ray spectra of several black
hole candidates that exhibit peaks at frequencies in a low integer ratio
could also potentially be associated with the orbital resonances
\cite{2006ARA&A..44...49R, 2004ApJ...606.1098S}. To aid the identification of astrophysical phenomena that originate from orbital resonances we fully characterize the region of parameter space where resonant effects occur. We present a number of easily evaluated formulae demonstrating the spin and eccentricity dependence of resonances and build an intuitive understanding for the inclination dependence. 

\emph{The resonance condition.}
Geodesic motion in the Kerr spacetime with spin parameter $a$, is integrable. The energy $E$, azimuthal angular momentum $L_z$ and Carter constant $Q$ fully specify the trajectory of a particle with rest mass $\mu$ \cite{Carter1968}.  
An equivalent description of the trajectory can be found 
using Kepler-type variables that are directly related to the orbit's geometry \cite{Schmidt}: its semi-latus rectum, eccentricity, and the cosine of the maximum orbital inclination, $\{p, e, z_-\}$.
For a generic bound orbit expressed in 
Boyer-Lindquist coordinates $(t, r, \theta, \phi)$, the radial motion
oscillates between the apastron, \mbox{$r_1=p/(1-e)$}, and the periastron, \mbox{$r_2=p/(1+e)$}, with a frequency $\omega_r$. 
The longitudinal motion oscillates about the equatorial plane with a frequency $\omega_\theta$, sampling the angles \mbox{$-\theta_*\leq \theta \leq \theta_*$},  we define \mbox{$z_- = \cos \theta^*$}.

Resonances occur for parameter values on a two-dimensional surface in $\{p$, $e$, $z_-\}$ space determined by the resonance condition \cite{Schmidt, ourpaper}
\begin{equation}
\frac{n}{m} = \frac{\omega_r}{\omega_{\theta}}= \left.\left( \int_{-z_-}^{z_-} \frac{  dz }{\sqrt{\Theta(z)}} \right)\middle/\left( \int_{r_2}^{r_1} \frac{dr}{\sqrt{R(r)}} \right),\right.  \label{RESCOND}
\end{equation}
where the functions $R$ and $\Theta$ 
can be factored as
\begin{align}
R&=-\beta^2(r-r_1)(r-r_2)(r-r_3)(r-r_4) \label{Rpotfac}\\
\Theta&=a^2 \beta^2 (z^2 - z_-^2)(z^2-z_+^2).         \label{rthpotfac}
\end{align}
Here, $\beta^2 = (\mu^2-E^2) $ and the roots obey $r_1\geq r_2 \geq r_3 \geq r_4$ and  $z_+\geq z_-$. 
  Evaluating the right-hand side of~\eqref{RESCOND} and extracting the physics of the resonant surfaces is complicated by the fact that the roots $r_3, r_4$ and $z_+$ are implicit functions of $\{p, e, z_-\}$.  The resonance condition \eqref{RESCOND} can be expressed in its most symmetric form using Carlson's integrals
\cite{1995NuAlg..10...13C}  
which allows us to obtain several useful analytic results in various limits \cite{ourpaper}. A rapidly convergent semi-analytical scheme for finding these surfaces in general can be constructed using these integrals \cite{ourpaper}.

\begin{figure}
\includegraphics[width = \columnwidth]
{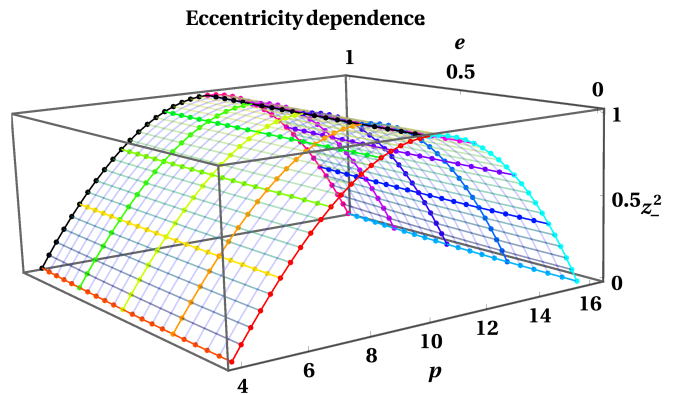}
\caption{\footnotesize{(Color online) The location of the $2/3$ resonance in $\{p, e, z_-\}$ parameter space.
The  
arch shape is typical for all resonances at fixed spin and depends very weakly on eccentricity.  
A maximum value of $z_-^2=1$ is reached at $p=p_{\rm polar}\sim p^*=10.8$. 
For a given $e$, the maximum (retrograde) and minimum (prograde) values of $p$ occur on the equatorial plane  $z_-=0$ and are indicated in blue (right) and red (left) respectively.
\label{res0304e}}}
\end{figure}
\emph{Features of resonance surfaces.} The $2/3$ resonance surface in $\{p,e,z_-\}$ space for a maximally spinning black hole is illustrated in Fig.~\ref{res0304e}. For a given spin, all resonant surfaces display the same qualitative  eccentricity and inclination dependence.
The resonance surface has the shape of an inverted $'U'$ arch that is weakly dependent on eccentricity and attains a maximum inclination of $z_-^2=1$ at  \mbox{$p=p_{\rm polar}$}, indicated by a black line in  Fig.~\ref{res0304e}.
For smaller inclination, $z_-^2<1$ and fixed eccentricity,  
the two possible values of $p$ on the resonant surface correspond to  prograde, $p_{+}<p_{\rm polar}$, closer to the black hole and retrograde, $p_{-}>p_{\rm polar}$, resonant orbits respectively. 
The $p_{\pm}$ subscript identifies $ {\rm sgn}(a L_z) =\pm 1$. 
As $z_-$ decreases
the distance $(p_{-}-p_{+})$ monotonically increases to its maximum value on the equatorial plane.

\begin{figure}
\includegraphics[width = \columnwidth]
{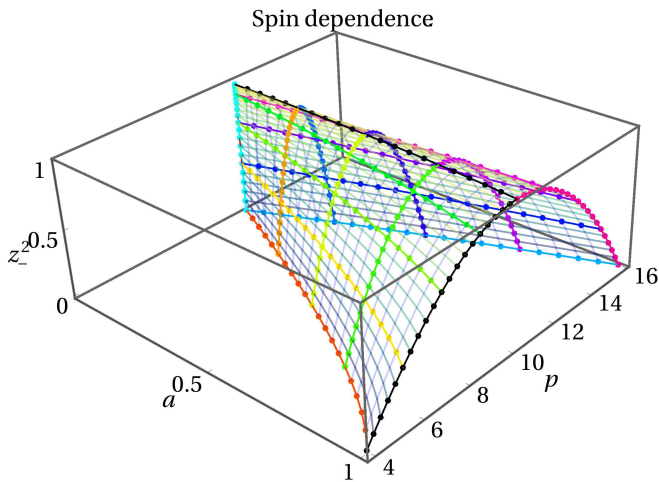}
\caption{\footnotesize{(Color online) Spin dependence for the  $2/3$ resonance with fixed \mbox{ $e=0.5$}.
Observe that as $a\rightarrow 0$ the arch pinches off to a line at
 $p=p_{\rm polar}$,  indicating that
resonances in this limit are independent of inclination. The maximum arch width occurs for a maximally spinning black hole  $a=1$.
\label{res0304ae0}
}}
\end{figure}

Since the basic features of a resonance for a given spin parameter $a$ depend weakly on eccentricity,
 we will now choose a representative eccentricity and study the spin dependence. Fig.~\ref{res0304ae0} shows the $2/3$ resonance surface for an eccentricity of $e=0.5$ as a function of black hole spin and $p$.
Here we see that the arch-width exhibits a strong spin dependence, with its inverted $'U'$ profile pinching off to a single column $'I'$ profile at  $p=p_{\rm polar}$ when $a\rightarrow 0$. This indicates that resonances in Schwarzschild spacetime are independent of inclination because the longitudinal frequency is just the same as the $\phi$-frequency in this case. As the spin increases the opening angle of the arch increases to a maximum arch width for $a=1$. The result is a $'V'$-shaped footprint of the arch in the $p$, $a$ plane. 
 As inclination increases, the prograde and retrograde branches of the arch approach $p_{\rm polar}$ 
and the $'V'$ profile narrows from its largest opening angle on the equatorial plane to a line for $z_-=1$. 
\begin{figure}[t]
\includegraphics[width = \columnwidth]{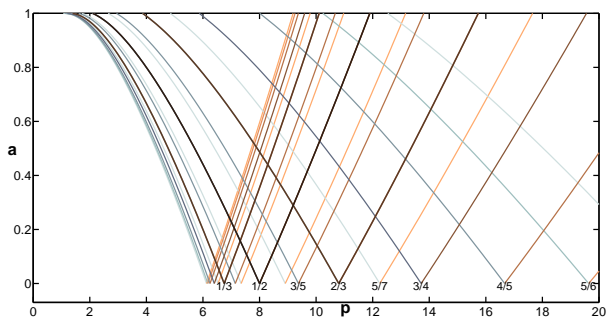}
\caption{\footnotesize{(Color online) The location of low order resonances ($m\leq 7$) for $e=z_-=0$ as a function of $a$ and  $p$.
 For $a=0$ the prograde and retrograde branches are degenerate at $p=p^*$.
Each vertex is labeled by the rational ratio $n/m$ and darker colors indicate lower order resonances. 
As the spin increases the retrograde branch leans right (copper tinge) and the prograde branch leans left (blue tinge).
\label{fig:spine0}}}
\end{figure}

A key result derived in \cite{ourpaper} is that
an exact analytic solution for the $'V'$ profile of circular equatorial orbits can be found. Since the $'U'$ profile is only weakly dependent on eccentricity this allows us to benchmark the location of the $n/m$ resonance for a black hole of arbitrary spin.  When  $e=z_-=0$,  Eq.~\eqref{RESCOND} is equivalent to  
\begin{align}
\left[p (p-p^*)-a^2 (p^*-3)\right]^2-4 a^2 p (p^*-2)^2=0.
\label{polyp}
\end{align}
where $p^*$  identifies the resonance we are considering and is related to the ratio $n/m$  by
\begin{align}
p^*=\frac{6}{1-(n/m)^2}.
\end{align}
For non spinning black holes,  $p=p^*$ is solution to Eq.~\eqref{polyp} and determines the position of the $'I'$ column in the $'U'$- $'I'$ transition for circular orbits.  
The value of $p^*$ sets the general mean radius in physical space about which all the interesting features associated with the $n/m$ resonance occur. Numerical values of $p^*$ for several low order resonances are given in Table.~\ref{TabRes}.
For spinning black holes, the largest two roots of Eq.\eqref{polyp} yield the $'V'$ profile on the equatorial plane. This profile is plotted for several low-order resonances in Fig.~\ref{fig:spine0}.
\begin{table}
\begin{tabular}{cccccc} \hline
Res. & Location  &  Period $T$ & Galactic &center:& Sgr A* \\\hline
$n/m$& $p^*$  $\left[GM/c^2  \right]$ & $\left[GM/c^3\right]$ & $p^*$  [$\mu {\rm a s}$] & $T$ [${\rm min}$] & $f$ [$10^{-4}{\rm Hz}$]  \\ \hline \hline
 ISCO& 6  &92.3 & 30.6  & 32.7 & 5.10 \\ 
 1/2 &  8  &  142.1 &  40.9  &  50.3  & 3.31  \\ 
 1/3 &  6.8  &  110.2 &  34.5&  39.0  &  4.27  \\
 \bf 2/3 & \bf 10.8  & \bf 223.0 & \bf 55.2  & \bf 78.9  & \bf 2.11  \\ 
 1/4 & 6.4  & 101.7 & 32.7 & 36.0  & 4.63  \\
 3/4 & 13.7  & 319.1 & 70.1  & 112.9   & 1.48  \\ 
 1/5 &  6.3  & 98.2 & 31.9     & 34.7   & 4.80  \\
 2/5 &  7.1  & 119.9 & 36.5 & 42.4   & 3.93  \\
 3/5 &  9.4  & 180.4 & 47.9  & 63.8   & 2.61  \\
 4/5 &  16.7 & 427.5 & 85.1 & 151.3   & 1.10  \\ \hline
\hline
\end{tabular}
\caption{\footnotesize{Time and length scales associated with low-order resonances. The values are for the $e=a=z_-=0$  vertexes seen in Fig.~\ref{fig:spine0}, both in dimensionless units scaled by the black hole's mass $M$ and in physical units for the special case of the Galactic center, Sgr A*. 
The values are calculated from $p^*=6/[1-(n/m)^2]$ and $T=2\pi p^{*3/2}$.
\label{TabRes}}
}
\end{table}
The maximum splitting of the retrograde and prograde branches occurs when $a=1$  in which case the relevant roots of Eq.~\eqref{polyp} are
 $p_{\pm} = p^*-1\mp2\sqrt{p^*-2}.$ For low spin values, the series expansion
\begin{align}
p_\mp=p^*\pm\frac{2 a (p^*-2)}{\sqrt{p^*}}-\frac{a^2 \left(p^{*2}-5
   p^*+8\right)}{p^{*2}} +O\left(a^3\right) \label{spinEQ}
\end{align}
is useful for making astrophysical estimates.
The weak eccentricity dependence of the $'U'$ profile for $a=0$ is
\begin{align}
\frac{p}{p^*}=1+\frac{e^2}{4(p^*-6)} - \frac{e^4(4p^*-17)}{64(p^*-6)^3} + O(e^6).
\label{pandecc}
\end{align} 
Observe that as the resonant surfaces approach the innermost stable circular orbit ($p=6$) the effects of eccentricity become increasingly important.
\begin{figure}
\includegraphics[width = \columnwidth]{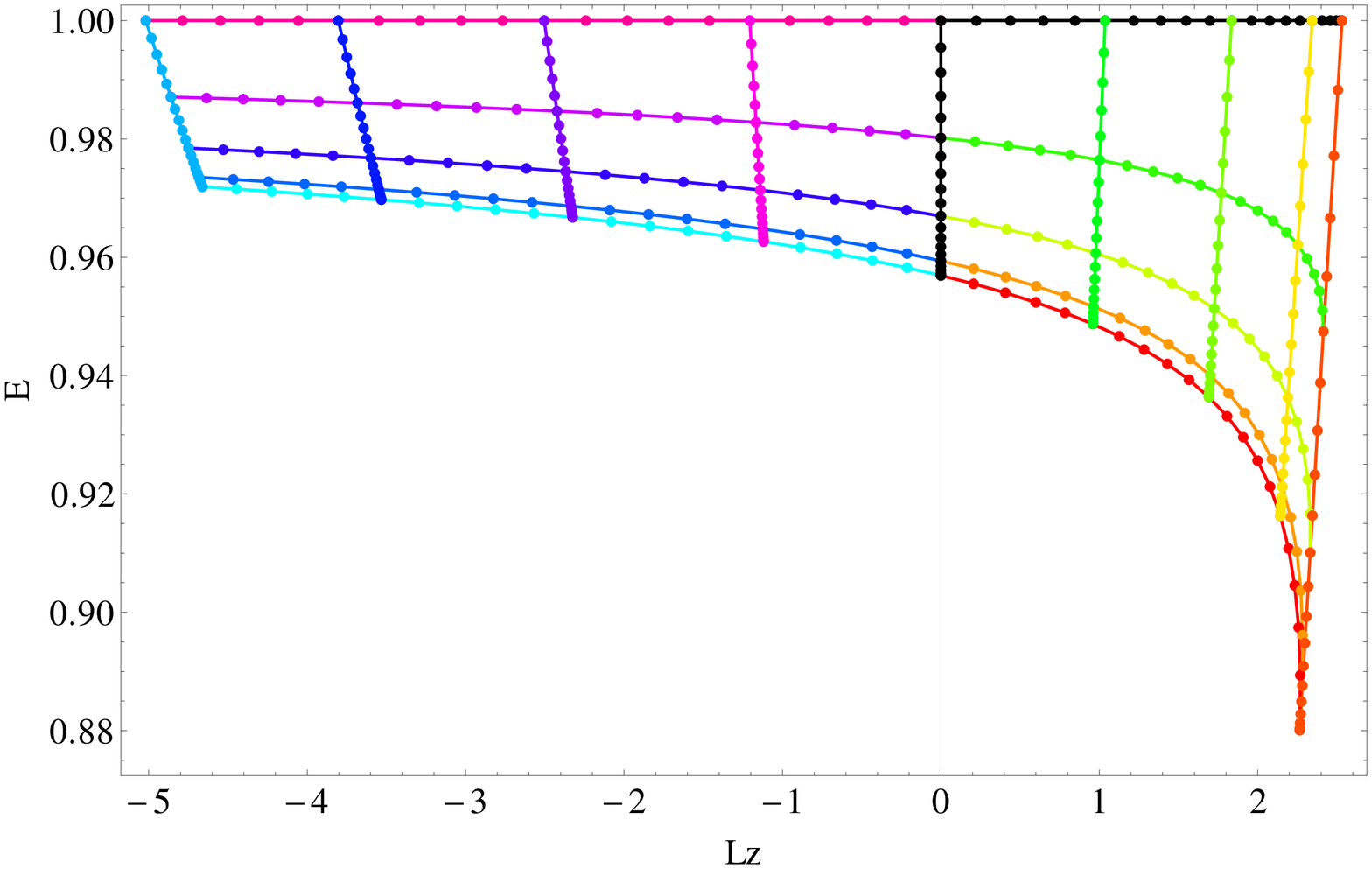}
\caption{\footnotesize{(Color online) The $2/3$ resonant surface for  $a=1$ projected onto $E$, $L_z$ coordinates.
 Lines indicate constant values of \mbox{$z_-^2 =\{ 0, \sin\frac{\pi}{8}, 1/\sqrt{2}, \cos\frac{\pi}{8},1\}$}, and \mbox{$e=\{0,\frac{1}{4},\frac{1}{2}, \frac{3}{4}, 1\}$} corresponding to the colored lines in Fig.~\ref{res0304e}. The $L_z=0$ line corresponds to the $z_-=1$ line in Fig.~\ref{res0304e}, the sharp cusp for $L_z>0$ to the prograde $e= z_-=0$ point on the arch (left front), the lower corner point for $L_z<0$ to the arch's retrograde  $e= z_-=0$ point (right front) and the $E= 1$ line to $e= 1$ for the arch.  
The large asymmetry across the $L_z = 0$ plane is as result of the high spin value.
\label{eplELZ2}}}
\end{figure}

\emph{Astrophysical implications.} 
Generally resonances can have either a capturing or a destabilizing effect on particles that enter their region of influence. We expect  Kerr resonances and the $2/3$ resonance in particular to have a stabilizing influence under perturbations\cite{ourpaper}. For particles that are light enough, 
 entering a resonance zone can temporarily halt the radiation-reaction driven orbital evolution because of an enhanced exchange of energy and angular momentum
from the black hole to the particle orbit. 
If a particle becomes captured by a resonance its orbital parameters are expected to change within the resonant surface \cite{Chicone2}
and for generic perturbations the orbit should simply evolve to the lowest energy state \cite{Chicone}. In Fig.~\ref{eplELZ2} we show the orbital energy and azimuthal angular momentum associated with the resonance surface that was depicted in $\{p,e,z_-\}$ space in Fig.~\ref{res0304e}. The lowest orbital energy state for the given resonance occurs in the lower right-hand corner of Fig ~\ref{eplELZ2} corresponding to prograde circular equatorial orbits. The migration of resonantly capture particles to circular equatorial configurations could result in a cohesive resonant structure where the ejection of particles due to collisional interactions is minimal. This should result in a distinctive imprint of density inhomogeneities on any thin disk surrounding a black hole similar to the structure imprinted on Saturn's rings \cite{rings, solarsystem}. 
Unlike the rings of Saturn whose structure is largely stationary, 
the black hole rings would be dynamical because the effect of radiation reaction dissipation will alter the resonant structures and enable escape from resonance \cite{Robinson, Chicone2}.
Once a trapped over-density becomes sufficiently large the radiation reaction force will dominate over the resonance's trapping potential 
and the ring will break, depositing an over-density of matter on the next ring in the disk.
Thus in one related catastrophic event the whole ring structure will re-adjust. 
The radiation emitted in the process is likely to be modulated with the characteristic frequencies associated with the collapsing resonance bands. The X-ray spectrum of a black hole candidate shows QPOs at pairs of frequencies in a 3:5 ratio in addition to the 2:3 ratio also observed in other systems \cite{2006ARA&A..44...49R}. Observing a $3:5$ frequency ratio is unexpected; from a dynamical systems perspective it is expected to be weaker than the 3:4 resonance. A possible explanation for its occurrence is that the physical location for the 3/5 resonance ($\sim$ 9.4M) occurs just inside the 2/3 resonance ($\sim$ 10.8 M) for all spin values as shown in Table \ref{TabRes} and  Fig.~\ref{fig:spine0}.
Matter from a disruption event of the 2/3 resonance structure could collide with even a tenuous over-density of matter at the 3/5 resonance location due to their close physical proximity and stimulate photon emission.

The results presented in this paper may provide a robust method of determining the black hole's spin given observational evidence from more than one resonance. 
Recent monitoring of Sgr A* with the 1.3mm VLBI showed time-variable
structures on scales $\sim 4 \, R_s$ \cite{2008Natur.455...78D, 2011ApJ...727L..36F}. The physical origin of this structure is not yet clear, but the scale is similar to that of the low-order resonances given in Table~\ref{TabRes}.
For argument's sake suppose that the origin of the structure at $\sim 4 R_S=8M$  is due to the $2/3$ resonance that  is displaced from its non-spinning position listed in Table~\ref{TabRes} due to Sgr A*'s spin. 
 On astrophysical grounds the $2/3$ resonance is likely to have the greatest probability of being directly observable \cite{ourpaper}.
Using Eq. \eqref{spinEQ} the spin displacement of the prograde 2/3 resonant surface is $p_+/M=10.8 - 5.36 a$, and comparing to the observed structure indicates that Sgr A* has spin $a=0.5$. The plausibility of identifying this structure with the 2/3 resonance could be confirmed 
if characteristic timescales of slightly less than an hour are associated with the variability and a 2:3 ratio in observed frequencies is discovered.  Note that once the spin is determined Eq. \eqref{spinEQ} predicts the location of the other resonances. As the sensitivity of the VLBI measurements increase and more of the horizon scales are resolved the observation of further resonances provides an independent check on the spin determined using the $2/3$ resonance.

Long-term monitoring of time of arrival signals from a pulsar with orbital period of a few months with the Square Kilometer array 
could determine the mass, spin and quadrupole moment of Sgr A* to a precision of $\lesssim 10^{-2}$  \cite{Liuetal12}, providing an exciting prospect for a definitive test of the no hair theorems.
A corollary  of the results in this paper is that 
orbits with periods of order months are sufficiently far from the low-order resonances that the KAM theorem guarantees the region to be effectively free of stochastic motion. Frequency drifts computed using perturbative methods based on averaging as done in \cite{2011CQGra..28v5029S} thus accurately describe the physical system.
Tracking the trajectory of a pulsar in the  region
 $50 R_s <r <1000 R_s$ 
should build up an accurate map of the central object's gravitational potential. 

From Table~\ref{TabRes} we observe that future space-based gravitational wave detectors sensitive to frequencies
\mbox{$\sim 10^{-4}-10^{-1}$} Hz will directly probe resonant dynamics. This is an  exciting possibility but it underscores the necessity of carefully modeling and incorporating resonant effects in the search templates. The resonance induced phase shift could potentially make careful parameter estimation difficult. If the central object is a non-Kerr Black hole the possible onset of geodesic chaos will occur first in these regions and further complicate the analysis. 
Further numerical investigation to fully quantify these effects for all $E$ and $L_z$ values, given in   Fig.~\ref{eplELZ2} and \cite{ourpaper}, associated with resonance surfaces is required.

\emph{Conclusion.}
We have explored the basic properties of resonant surfaces associated with radial and longitudinal motion around a Kerr black hole. A few simple expressions to quantify resonant effects 
in astrophysical systems with strong gravitational fields have been provided. We have suggested a resonance-based method for determining the spins of black holes. Observations of QPOs, gravitational wave emission from resonant transits and radio maps of Sgr A* at event horizon scales will in the near future provide a powerful observational toolkit for probing resonant phenomena.

\emph{Acknowledgments.} This work was supported in part by NSF Grants PHY-0903631 and PHY-1208881, and the Maryland Center for Fundamental Physics.

\bibliography{BholesNemadon,ResonanceRefs}

\begin{thebibliography}{29}
\expandafter\ifx\csname natexlab\endcsname\relax\def\natexlab#1{#1}\fi
\expandafter\ifx\csname bibnamefont\endcsname\relax
  \def\bibnamefont#1{#1}\fi
\expandafter\ifx\csname bibfnamefont\endcsname\relax
  \def\bibfnamefont#1{#1}\fi
\expandafter\ifx\csname citenamefont\endcsname\relax
  \def\citenamefont#1{#1}\fi
\expandafter\ifx\csname url\endcsname\relax
  \def\url#1{\texttt{#1}}\fi
\expandafter\ifx\csname urlprefix\endcsname\relax\def\urlprefix{URL }\fi
\providecommand{\bibinfo}[2]{#2}
\providecommand{\eprint}[2][]{\url{#2}}

\bibitem[{\citenamefont{{Murray} and {Holman}}(2001)}]{2001Natur.410..773M}
\bibinfo{author}{\bibfnamefont{N.}~\bibnamefont{{Murray}}} \bibnamefont{and}
  \bibinfo{author}{\bibfnamefont{M.}~\bibnamefont{{Holman}}},
  \bibinfo{journal}{Nature} \textbf{\bibinfo{volume}{410}},
  \bibinfo{pages}{773} (\bibinfo{year}{2001}).

\bibitem[{\citenamefont{H\"{a}nninen}(2001)}]{Saturn}
\bibinfo{author}{\bibfnamefont{J.}~\bibnamefont{H\"{a}nninen}},
  \emph{\bibinfo{title}{Formation of Narrow Ringlets in Saturn’s Rings}},
  vol. \bibinfo{volume}{564} of \emph{\bibinfo{series}{Lecture Notes in
  Physics}} (\bibinfo{publisher}{Springer Berlin Heidelberg},
  \bibinfo{year}{2001}).

\bibitem[{\citenamefont{Dermott and Murray}(1981)}]{Asteroid}
\bibinfo{author}{\bibfnamefont{S.~F.} \bibnamefont{Dermott}} \bibnamefont{and}
  \bibinfo{author}{\bibfnamefont{C.~D.} \bibnamefont{Murray}},
  \bibinfo{journal}{Nature} \textbf{\bibinfo{volume}{290}},
  \bibinfo{pages}{664} (\bibinfo{year}{1981}).

\bibitem[{\citenamefont{Schmidt}(2002)}]{Schmidt}
\bibinfo{author}{\bibfnamefont{W.}~\bibnamefont{Schmidt}},
  \bibinfo{journal}{Class. Quantum Grav.} \textbf{\bibinfo{volume}{19}},
  \bibinfo{pages}{2743} (\bibinfo{year}{2002}).

\bibitem[{\citenamefont{Semerák and Suková}(2012)}]{Semerák01102012}
\bibinfo{author}{\bibfnamefont{O.}~\bibnamefont{Semerák}} \bibnamefont{and}
  \bibinfo{author}{\bibfnamefont{P.}~\bibnamefont{Suková}},
  \bibinfo{journal}{Monthly Notices of the Royal Astronomical Society}
  \textbf{\bibinfo{volume}{425}}, \bibinfo{pages}{2455} (\bibinfo{year}{2012}).

\bibitem[{\citenamefont{{Contopoulos} et~al.}(2012)\citenamefont{{Contopoulos},
  {Harsoula}, and {Lukes-Gerakopoulos}}}]{Contopoulos2012}
\bibinfo{author}{\bibfnamefont{G.}~\bibnamefont{{Contopoulos}}},
  \bibinfo{author}{\bibfnamefont{M.}~\bibnamefont{{Harsoula}}},
  \bibnamefont{and}
  \bibinfo{author}{\bibfnamefont{G.}~\bibnamefont{{Lukes-Gerakopoulos}}},
  \bibinfo{journal}{Celestial Mechanics and Dynamical Astronomy}
  \textbf{\bibinfo{volume}{113}}, \bibinfo{pages}{255} (\bibinfo{year}{2012}),
  \eprint{1203.1010}.

\bibitem[{\citenamefont{Gair et~al.}(2008)\citenamefont{Gair, Li, and
  Mandel}}]{Gair}
\bibinfo{author}{\bibfnamefont{J.~R.} \bibnamefont{Gair}},
  \bibinfo{author}{\bibfnamefont{C.}~\bibnamefont{Li}}, \bibnamefont{and}
  \bibinfo{author}{\bibfnamefont{I.}~\bibnamefont{Mandel}},
  \bibinfo{journal}{Phys.\ Rev.\ D} \textbf{\bibinfo{volume}{77}},
  \bibinfo{pages}{024035} (\bibinfo{year}{2008}).

\bibitem[{\citenamefont{Brink}(2008)}]{JdB1}
\bibinfo{author}{\bibfnamefont{J.}~\bibnamefont{Brink}},
  \bibinfo{journal}{Phys.\ Rev.\ D} \textbf{\bibinfo{volume}{78}},
  \bibinfo{eid}{102002} (\bibinfo{year}{2008}).

\bibitem[{\citenamefont{Poisson et~al.}(2011)\citenamefont{Poisson, Pound, and
  Vega}}]{lrr-2011-7}
\bibinfo{author}{\bibfnamefont{E.}~\bibnamefont{Poisson}},
  \bibinfo{author}{\bibfnamefont{A.}~\bibnamefont{Pound}}, \bibnamefont{and}
  \bibinfo{author}{\bibfnamefont{I.}~\bibnamefont{Vega}},
  \bibinfo{journal}{Living Reviews in Relativity} \textbf{\bibinfo{volume}{14}}
  (\bibinfo{year}{2011}).

\bibitem[{\citenamefont{{Hartl}}(2003)}]{spin2}
\bibinfo{author}{\bibfnamefont{M.~D.} \bibnamefont{{Hartl}}},
  \bibinfo{journal}{\prd} \textbf{\bibinfo{volume}{67}}, \bibinfo{eid}{104023}
  (\bibinfo{year}{2003}).

\bibitem[{\citenamefont{Arnold}(1963)}]{Arnold1963a}
\bibinfo{author}{\bibfnamefont{V.~I.} \bibnamefont{Arnold}},
  \bibinfo{journal}{Russian Math. Survey} \textbf{\bibinfo{volume}{18}},
  \bibinfo{pages}{9} (\bibinfo{year}{1963}).

\bibitem[{\citenamefont{Moser}(1973)}]{Moser1973}
\bibinfo{author}{\bibfnamefont{J.}~\bibnamefont{Moser}},
  \emph{\bibinfo{title}{Stable and Random Motions in Dynamical Systems}}
  (\bibinfo{publisher}{Princeton, NJ : Princeton University Press},
  \bibinfo{year}{1973}).

\bibitem[{\citenamefont{{Doeleman}}(2010)}]{eventhorizontel}
\bibinfo{author}{\bibfnamefont{S.}~\bibnamefont{{Doeleman}}}, in
  \emph{\bibinfo{booktitle}{10th European VLBI Network Symposium and EVN Users
  Meeting: VLBI and the New Generation of Radio Arrays}}
  (\bibinfo{year}{2010}).

\bibitem[{\citenamefont{{Flanagan} and {Hinderer}}(2012)}]{prl}
\bibinfo{author}{\bibfnamefont{{\'E}.~{\'E}.} \bibnamefont{{Flanagan}}}
  \bibnamefont{and}
  \bibinfo{author}{\bibfnamefont{T.}~\bibnamefont{{Hinderer}}},
  \bibinfo{journal}{PRL} \textbf{\bibinfo{volume}{109}}, \bibinfo{eid}{071102}
  (\bibinfo{year}{2012}).

\bibitem[{\citenamefont{{Flanagan} et~al.}(2012)\citenamefont{{Flanagan},
  {Hughes}, and {Ruangsri}}}]{uchopol}
\bibinfo{author}{\bibfnamefont{E.~E.} \bibnamefont{{Flanagan}}},
  \bibinfo{author}{\bibfnamefont{S.~A.} \bibnamefont{{Hughes}}},
  \bibnamefont{and}
  \bibinfo{author}{\bibfnamefont{U.}~\bibnamefont{{Ruangsri}}}
  (\bibinfo{year}{2012}), \eprint{arXiv:gr-qc:1208.3906}.

\bibitem[{\citenamefont{{Fish} et~al.}(2011)\citenamefont{{Fish}, {Doeleman},
  {Beaudoin}, {Blundell}, {Bolin}, {Bower}, {Chamberlin}, {Freund}, {Friberg},
  {Gurwell} et~al.}}]{2011ApJ...727L..36F}
\bibinfo{author}{\bibfnamefont{V.~L.} \bibnamefont{{Fish}}},
  \bibinfo{author}{\bibfnamefont{S.~S.} \bibnamefont{{Doeleman}}},
  \bibinfo{author}{\bibfnamefont{C.}~\bibnamefont{{Beaudoin}}},
  \bibinfo{author}{\bibfnamefont{R.}~\bibnamefont{{Blundell}}},
  \bibinfo{author}{\bibfnamefont{D.~E.} \bibnamefont{{Bolin}}},
  \bibinfo{author}{\bibfnamefont{G.~C.} \bibnamefont{{Bower}}},
  \bibinfo{author}{\bibfnamefont{R.}~\bibnamefont{{Chamberlin}}},
  \bibinfo{author}{\bibfnamefont{R.}~\bibnamefont{{Freund}}},
  \bibinfo{author}{\bibfnamefont{P.}~\bibnamefont{{Friberg}}},
  \bibinfo{author}{\bibfnamefont{M.~A.} \bibnamefont{{Gurwell}}},
  \bibnamefont{et~al.}, \bibinfo{journal}{\apj} \textbf{\bibinfo{volume}{727}},
  \bibinfo{eid}{L36} (\bibinfo{year}{2011}).

\bibitem[{\citenamefont{{Remillard} and
  {McClintock}}(2006)}]{2006ARA&A..44...49R}
\bibinfo{author}{\bibfnamefont{R.~A.} \bibnamefont{{Remillard}}}
  \bibnamefont{and} \bibinfo{author}{\bibfnamefont{J.~E.}
  \bibnamefont{{McClintock}}}, \bibinfo{journal}{Annual Review of Astronomy and
  Astrophysics} \textbf{\bibinfo{volume}{44}}, \bibinfo{pages}{49}
  (\bibinfo{year}{2006}).

\bibitem[{\citenamefont{{Schnittman} and
  {Bertschinger}}(2004)}]{2004ApJ...606.1098S}
\bibinfo{author}{\bibfnamefont{J.~D.} \bibnamefont{{Schnittman}}}
  \bibnamefont{and}
  \bibinfo{author}{\bibfnamefont{E.}~\bibnamefont{{Bertschinger}}},
  \bibinfo{journal}{\apj} \textbf{\bibinfo{volume}{606}}, \bibinfo{pages}{1098}
  (\bibinfo{year}{2004}).

\bibitem[{\citenamefont{Carter}(1968)}]{Carter1968}
\bibinfo{author}{\bibfnamefont{B.}~\bibnamefont{Carter}},
  \bibinfo{journal}{Physical Review} \textbf{\bibinfo{volume}{174}},
  \bibinfo{pages}{1559} (\bibinfo{year}{1968}).

\bibitem[{\citenamefont{{Brink} et~al.}(2013)\citenamefont{{Brink}, {Geyer},
  and {Hinderer}}}]{ourpaper}
\bibinfo{author}{\bibfnamefont{J.}~\bibnamefont{{Brink}}},
  \bibinfo{author}{\bibfnamefont{M.}~\bibnamefont{{Geyer}}}, \bibnamefont{and}
  \bibinfo{author}{\bibfnamefont{T.}~\bibnamefont{{Hinderer}}},
  \bibinfo{journal}{in preparation}  (\bibinfo{year}{2013}).

\bibitem[{\citenamefont{{Carlson}}(1995)}]{1995NuAlg..10...13C}
\bibinfo{author}{\bibfnamefont{B.}~\bibnamefont{{Carlson}}},
  \bibinfo{journal}{Numerical Algorithms} \textbf{\bibinfo{volume}{10}},
  \bibinfo{pages}{13} (\bibinfo{year}{1995}).

\bibitem[{\citenamefont{{Chicone} et~al.}(2000)\citenamefont{{Chicone},
  {Mashhoon}, and {Retzloff}}}]{Chicone2}
\bibinfo{author}{\bibfnamefont{C.}~\bibnamefont{{Chicone}}},
  \bibinfo{author}{\bibfnamefont{B.}~\bibnamefont{{Mashhoon}}},
  \bibnamefont{and} \bibinfo{author}{\bibfnamefont{D.~G.}
  \bibnamefont{{Retzloff}}}, \bibinfo{journal}{Journal of Physics A
  Mathematical General} \textbf{\bibinfo{volume}{33}}, \bibinfo{pages}{513}
  (\bibinfo{year}{2000}).

\bibitem[{\citenamefont{{Chicone} et~al.}(1999)\citenamefont{{Chicone},
  {Mashhoon}, and {Retzloff}}}]{Chicone}
\bibinfo{author}{\bibfnamefont{C.}~\bibnamefont{{Chicone}}},
  \bibinfo{author}{\bibfnamefont{B.}~\bibnamefont{{Mashhoon}}},
  \bibnamefont{and} \bibinfo{author}{\bibfnamefont{D.~G.}
  \bibnamefont{{Retzloff}}}, \bibinfo{journal}{Classical and Quantum Gravity}
  \textbf{\bibinfo{volume}{16}}, \bibinfo{pages}{507} (\bibinfo{year}{1999}).

\bibitem[{\citenamefont{{Goldreich} and {Tremaine}}(1982)}]{rings}
\bibinfo{author}{\bibfnamefont{P.}~\bibnamefont{{Goldreich}}} \bibnamefont{and}
  \bibinfo{author}{\bibfnamefont{S.}~\bibnamefont{{Tremaine}}},
  \bibinfo{journal}{Annual review of astronomy and astrophysics}
  \textbf{\bibinfo{volume}{20}}, \bibinfo{pages}{249} (\bibinfo{year}{1982}).

\bibitem[{\citenamefont{{Peale}}(1976)}]{solarsystem}
\bibinfo{author}{\bibfnamefont{S.~J.} \bibnamefont{{Peale}}},
  \bibinfo{journal}{Annual review of astronomy and astrophysics}
  \textbf{\bibinfo{volume}{14}}, \bibinfo{pages}{215} (\bibinfo{year}{1976}).

\bibitem[{\citenamefont{{Robinson}}(1983)}]{Robinson}
\bibinfo{author}{\bibfnamefont{C.}~\bibnamefont{{Robinson}}},
  \bibinfo{journal}{Lecture Notes in Mathematics, Berlin Springer Verlag}
  \textbf{\bibinfo{volume}{1007}}, \bibinfo{pages}{651} (\bibinfo{year}{1983}).

\bibitem[{\citenamefont{{Doeleman} et~al.}(2008)\citenamefont{{Doeleman},
  {Weintroub}, {Rogers}, {Plambeck}, {Freund}, {Tilanus}, {Friberg}, {Ziurys},
  {Moran}, {Corey} et~al.}}]{2008Natur.455...78D}
\bibinfo{author}{\bibfnamefont{S.~S.} \bibnamefont{{Doeleman}}},
  \bibinfo{author}{\bibfnamefont{J.}~\bibnamefont{{Weintroub}}},
  \bibinfo{author}{\bibfnamefont{A.~E.~E.} \bibnamefont{{Rogers}}},
  \bibinfo{author}{\bibfnamefont{R.}~\bibnamefont{{Plambeck}}},
  \bibinfo{author}{\bibfnamefont{R.}~\bibnamefont{{Freund}}},
  \bibinfo{author}{\bibfnamefont{R.~P.~J.} \bibnamefont{{Tilanus}}},
  \bibinfo{author}{\bibfnamefont{P.}~\bibnamefont{{Friberg}}},
  \bibinfo{author}{\bibfnamefont{L.~M.} \bibnamefont{{Ziurys}}},
  \bibinfo{author}{\bibfnamefont{J.~M.} \bibnamefont{{Moran}}},
  \bibinfo{author}{\bibfnamefont{B.}~\bibnamefont{{Corey}}},
  \bibnamefont{et~al.}, \bibinfo{journal}{\nat} \textbf{\bibinfo{volume}{455}},
  \bibinfo{pages}{78} (\bibinfo{year}{2008}).

\bibitem[{\citenamefont{{Liu} et~al.}(2012)\citenamefont{{Liu}, {Wex},
  {Kramer}, {Cordes}, and {Lazio}}}]{Liuetal12}
\bibinfo{author}{\bibfnamefont{K.}~\bibnamefont{{Liu}}},
  \bibinfo{author}{\bibfnamefont{N.}~\bibnamefont{{Wex}}},
  \bibinfo{author}{\bibfnamefont{M.}~\bibnamefont{{Kramer}}},
  \bibinfo{author}{\bibfnamefont{J.~M.} \bibnamefont{{Cordes}}},
  \bibnamefont{and} \bibinfo{author}{\bibfnamefont{T.~J.~W.}
  \bibnamefont{{Lazio}}}, \bibinfo{journal}{\apj}
  \textbf{\bibinfo{volume}{747}}, \bibinfo{eid}{1} (\bibinfo{year}{2012}).

\bibitem[{\citenamefont{{Sadeghian} and {Will}}(2011)}]{2011CQGra..28v5029S}
\bibinfo{author}{\bibfnamefont{L.}~\bibnamefont{{Sadeghian}}} \bibnamefont{and}
  \bibinfo{author}{\bibfnamefont{C.~M.} \bibnamefont{{Will}}},
  \bibinfo{journal}{Classical and Quantum Gravity}
  \textbf{\bibinfo{volume}{28}}, \bibinfo{pages}{225029}
  (\bibinfo{year}{2011}).

\end{thebibliography}

\end{document}